# Simulative Performance Analysis of an AD Function with Road Network Variation

**Daniel Becker[1*], Guido Küppers[1], Lutz Eckstein[1]**

1. Institute for Automotive Engineering (ika) RWTH Aachen University, Germany (daniel.becker@ika.rwth-aachen.de)

**Abstract**

Automated driving functions (ADFs) have become increasingly popular in recent years. However, their safety must be assured. Thus, the verification and validation of these functions is still an important open issue in research and development. To achieve this efficiently, scenario-based testing has been established as a valuable methodology among researchers, industry, as well as authorities. Simulations are a powerful way to test those scenarios reproducibly. In this paper, we propose a method to automatically test a set of scenarios in many variations. In contrast to related approaches, those variations are not applied to traffic participants around the ADF, but to the road network to show that parameters regarding the road topology also influence the performance of such an ADF. We present a continuous tool chain to set up scenarios, variate them, run simulations and finally, evaluate the performance with a set of key performance indicators (KPIs).

**Keywords:**

Automated Driving, Road Network Variation, Scenario-Based Testing, Simulation Maps

**Introduction**

Automated driving functions (ADFs) and advanced driver assistant systems (ADAS) have the potential to enhance the road safety significantly since according to (Statistisches Bundesamt 2022) 88% of all accidents are caused by humans. One major challenge while releasing those functions to the public, and ultimately increasing road safety, is the safety assurance. As claimed by (Wachenfeld and Winner 2016), classical validation methods of ADAS, such as test drives on public roads, would require billions of test kilometers to argue that an ADF is safe enough to operate on public roads. To overcome these distances, scenario-based testing has been identified as a suitable solution (Zlocki et al. 2015). The idea is to only test situations which are relevant for the safety of an ADF, then structure those scenarios and roll them out systematically to ensure the function's safety. This approach is pursued in the German "Pegasus project family"[1] where an overall methodology for the verification and validation for automated vehicles is developed. Recent research has its focus mainly on the systematic variation of surrounding dynamic objects of an ADF, e.g., (Weber et al. 2019) describe a framework to structure possible situations that can arise on motorways between an ADF and other traffic participants. However, the design and geometry of road networks is an important safety factor as well which is for example discussed in (Berger 2016). This is why the focus of this work is on the static part of scenarios to show the impact of road network design on the safety level of an ADF.

---

[1] https://pegasus-family.de/

One of the key tools used in the development and testing of automated driving systems is simulation software because it allows developers to test their systems in a controlled environment before deploying them in real vehicles on test tracks or public roads. This can help ensure the safety and reliability of the automated driving system, as well as help identify and fix any potential issues before the system is deployed. There are several commonly used simulation software platforms in the field of automated driving, like *CarMaker*, *Virtual Test Drive* (*VTD*) or *Carla*. *CarMaker* and *VTD* are commercial simulation software developed by IPG Automotive and Hexagon (formerly Vires), respectively. They are widely used in the automotive industry for the development and testing of vehicle dynamics and control systems. *Carla* is an open-source platform that provides a high-fidelity simulation environment for the development and testing of autonomous vehicles and comes with a powerful API for developing and testing algorithms as well as an interface to the Robot Operating System (ROS).

ROS is an open-source project providing a range of tools and libraries for developing and testing robotic systems. In the context of automated driving, ROS is often used to develop and test algorithms for autonomous vehicles. One of the major advantages is the principle of *Publisher* and *Subscriber* enabling a straightforward modular software architecture so that the developed modules can be evaluated with real-world sensor data as well as simulated inputs.

This paper describes a method to confront a prototypical ADF with automatically generated road networks followed by an evaluation of the performed scenarios in a simulation environment. The remainder of this paper is structured as follows. First, a brief overview on related work is given to make sure the reader can follow the presented concept which is described in section Methodology and then detailed in the following sections about scenario generation and test execution. Subsequently, some simulation examples are evaluated and discussed, followed by a conclusion and an outlook on possible further working steps.

The main contributions to the state-of-the-art are:
- We present a framework that allows the automated testing of a set of logical scenarios instantiated with an arbitrary number of concrete scenarios.
- We combine available tools and concepts to achieve an approach that can be used with various tools and models.
- We evaluate a prototypical automated driving function in IPG *CarMaker* according to a set of KPIs to show the influence of the road topology on the ADFs performance.

**Related Work**

The presented approach is based on previous work of the authors as described in (Becker et al. 2020) and (Becker et al. 2022) where the process of formulating road networks in a straightforward manner is introduced as well as the automated generation of simulation maps in the OpenDRIVE standard. This section gives a brief overview on related work regarding scenario concepts and automated driving platforms.

*Scenario Concepts*

Within the scenario-based testing approach, there are several concepts and formats which are used. This section introduces some of them to have a common basis for the developed methodology. To efficiently specify a scenario, it seems useful to structure scenarios which has been done by (Bagschik et al. 2018) or (Bock et al. 2018) in the course of the Pegasus research project. The result was a five-, respectively six-layer model where each layer describes a part of the environment. The first layer defines details about the road layout (i.e., road network and traffic guidance), whereas the second layer covers roadside elements such as buildings or trees. Temporal modifications of the first and second layer are defined in the third layer and dynamic objects such as vehicles or pedestrians are described within layer four. Any weather and digital information is stored in layers five and six, respectively. Since the Pegasus project solely focused on motorway, the model was further developed and systematically defined in the VVMethoden project for urban traffic. Figure 1 illustrates the described layers and a detailed elaboration can be found in (Scholtes et al. 2021).

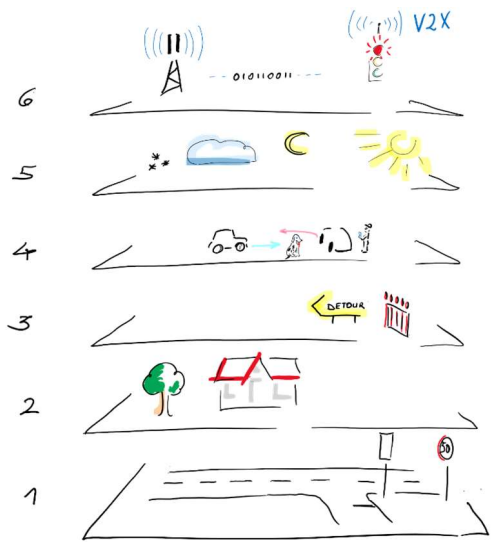

**Figure 1: The 6-Layer-Modell according to (Scholtes et al. 2021)**

Besides a common understanding on how to structure a scenario, it is important that those can be defined and stored in standardized formats. For the road network there exist several formats which are either required for a given simulation environment or have other benefits when using them. E.g., the ASAM OpenDRIVE format, maintained and developed by ASAM e.V., has become a widely used standard to exchange a detailed model of the simulation map. However, OpenDRIVE consists of continuous formulations for the road course which is exact and efficient but may be disadvantageous when used in a driving function which requires discrete (fine sampled) points for its trajectory planning. For this purpose the Lanelet2 format has become a valuable format, presented by (Poggenhans et al. 2018) and is inspired by OpenStreetMap[2]. The format comes with a C++ library to handle HD map data efficiently. Closely related to Lanlet2 is the CommonRoad format which makes some modification to the Lanelet2 format in terms of the road network. However, CommonRoad also describes the remaining scenario parts (cf. (Maierhofer et al. 2021)). In addition, the authors presented a toolbox which can convert

---

[2] https://www.openstreetmap.org

OpenStreetMap, Lanelet2, OpenDRIVE, and SUMO into their format as well as an export the road network to Lanelet2. Another format which is used in *IPG CarMaker* is their own *road5* format (cf. (IPG Automotive Group 2020)).

As stated in the six-layer model, besides the road network a scenario consists of more aspects. For this, the ASAM OpenSCENARIO standard has evolved as a popular format. Its primary use-case is to describe complex, synchronized maneuvers that involve multiple entities like vehicles, pedestrians and other traffic participants (ASAM e. V. 2022).

*Automated Driving Platforms*

As described in the introduction, scenario-based testing is a suitable solution for verification and validation of automated driving functions. To be able to perform scenario-based testing, or like in this case, perform research on these methods an ADF needs to be present. While most of these software stacks in industry, but also in research are proprietary there also exist open-source frameworks that provide a strong foundation for the development of ADF like Apollo[3] and Autoware[4]. The proposed method utilized ika's ADF as the so-called system under test (SUT). This ADF is a prototype function, which has been developed both in simulation and by testing in a test vehicle according to (Kueppers et al. 2022). The ADF is implemented in C++ using ROS. This allows a modular functional architecture which gives the opportunity to test individual sub-functionalities of the system in a component-wise manner using simulation tools.

**Methodology**

We propose a framework that allows to test variations in the road network in connection with a prototypical ADF and the commercial simulation tool IPG *CarMaker*. The workflow is outlined in Figure 2 and can be described as follows. First, a logical description of the road network according to (Becker et al. 2020) is formulated. Then, arbitrary parameters can be randomly variated to generate a set of OpenDRIVE maps which are then converted into the required formats of the used simulation modules. Consequently, the prototypical ADF is set up in its environment and connected to the simulation software. Subsequently, the simulations with each generated simulation map are performed and finally the results are evaluated according to a set of KPIs. In the following sections, the steps just presented will be detailed.

**Scenario Generation**

In this section, the blue boxes shown in Figure 2 are subsequently described. For a first application of the methodology, two simple road network templates are used which are briefly documented in the following section. Next, the parameters which are variated are introduced and finally, the generation of concrete instances of the logical scenario will be explained in more detail.

---

[3] https://developer.apollo.auto/

[4] https://www.autoware.org/

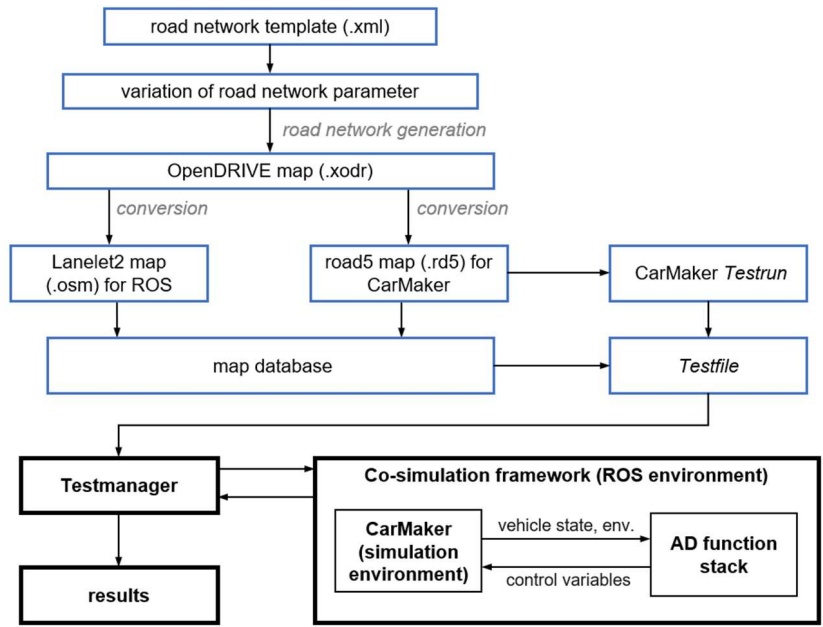

**Figure 2: Overview of the implemented tool chain.**

*Logical Scenario*

To show that the developed methodology meets its goal, two rather straightforward road networks are chosen. Namely, a road with two lanes in each driving direction describing a straight line, followed by a curve. The second map is a T-junction. Examples of both maps are shown in Figure 3. Both scenarios are solely defined in their plan view and elevation as well as hyper elevation are not considered at this point. The road networks are written in an xml syntax which was presented in (Becker et al. 2020) and is inspired by OpenDRIVE but with reduced redundancies and which allows a more logical formulation of a road network.

Besides the road network, both scenarios consist of a system-under-test (SUT) which is placed at the start of the road for the single road (i.e., onto the straight line on left end of the curve in Figure 3 a). Its goal is the reach the end of the road and to pass the curve. For the T-junction, the SUT is placed at the arm of the "T" (i.e., the road shown the bottom of Figure 3 b) and its desired position is the right arm of the T-junction. The designated maneuver is a right turn. Other traffic participants are not considered in the present scenario definition since in this first step only the effect of road network parameters shall be analyzed.

Right now, for the logical description of the scenarios no dedicated language is used, but the concrete IPG *CarMaker* test files are manipulated with help of Python scripts. The process is described during this paper (cf. section Concrete Scenario).

Simulative Performance Analysis of an AD Function with Road Network VariationSimulative Performance Analysis of an AD Function with Road Network Variation

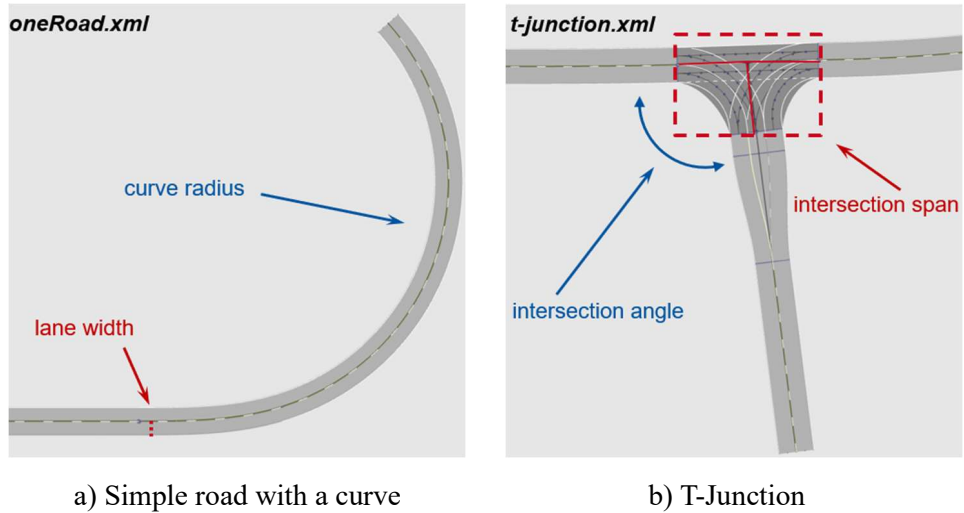

a) Simple road with a curve     b) T-Junction

**Figure 3: Road network structure of the two logical scenarios used for the analysis.**

*Parameter Selection and Sampling*

The conversion of logical to concrete scenarios is an important step in the presented tool chain. That is why before generating concrete scenarios, it is important to select meaningful parameters and their ranges from which the concrete values are sampled. In case of the selected scenarios, two parameters were chosen in each case, which are indicated in Figure 3. For the curved road lane width and curve radius are modelled as stochastic variables and within the T-junction the intersection span (start of the turning lanes from the intersection's center) and the intersection angle are variated. Table 1 summarizes all parameters that are modeled as stochastic variables. Each parameter is carried out in three sets, respectively. Note that the remaining parameters are kept constant and only one parameter is varied for each test.

**Table 1: Parameters for the road network variation defined as normal distributions**

| Road Network | Parameter | Values | Remarks |
|---|---|---|---|
| Curved road | lane width | Set 1: $\mu = 3.0\ m, \sigma = 0.05\ m$<br>Set 2: $\mu = 3.25\ m, \sigma = 0.1\ m$<br>Set 3: $\mu = 3.75\ m, \sigma = 0.1\ m$ | Radius constant at $50\ m$ |
| Curved road | curve radius | Set 1: $\mu = 12.0\ m, \sigma = 0.5\ m$<br>Set 2: $\mu = 20.0\ m, \sigma = 1.0\ m$<br>Set 3: $\mu = 30.0\ m, \sigma = 1.0\ m$ | Lane width constant at $3.25\ m$ |
| T-Junction | intersection angle | Set 1: $\mu = 45\ deg, \sigma = 5.73\ deg$<br>Set 2: $\mu = 90\ deg, \sigma = 11.46\ deg$<br>Set 3: $\mu = 135\ deg, \sigma = 5.73\ deg$ | Intersection span constant at $15\ m$ |
| T-Junction | intersection span | Set 1: $\mu = 15.0\ m, \sigma = 1.0\ m$<br>Set 2: $\mu = 20.0\ m, \sigma = 1.0\ m$<br>Set 3: $\mu = 30.0\ m, \sigma = 1.0\ m$ | Intersection angle constant at $90\ deg$ |

The selected parameters and their ranges origin from road construction regulations such as (Baier 2007), and the analysis of real road networks, e.g., described in (Becker et al. 2022).

Simulative Performance Analysis of an AD Function with Road Network Variation

*Concrete Scenario*

In order to run simulations, the defined logical scenarios need to be instantiated as concrete scenarios. This is done in an automated process that uses some self-developed and given tools. Additionally, *CarMaker* "Testrun" files need to be manipulated since they do not support logical formulations. The process is structured as follows. First, the logical road networks are transformed into OpenDRIVE with help of a *RoadGeneration*[5] tool developed by ika. Those standardized OpenDRIVE files need to be converted into IPG's *road5* format which is done by their *roadutil* tool via command line. Additionally, the ADF requires a *Lanelet2* map. Fortunately, this can be done with the open-source tool *CommonRoad Scenario Designer* (former *opendrive2lanelet*)[6]. Once the concrete maps are generated, the "Testrun" file for *CarMaker* needs to be adjusted so that the correct *road5* map will be loaded and the according start and desired end point of the ADF is set correctly. This is done by a python script. All generated files are stored in a database where a "testmanager" selects and starts all defined scenarios (cf. Figure 2 for an overview of the workflow). Within the next section, the actual simulation test run setup in the ROS environment is described.

**Test Simulation Setup**

As described in the methodology, a co-simulation framework to test the ika ADF is implemented. As simulation-tool IPG *CarMaker* is used. Since the ADF bases on ROS, the *CarMaker* C-Code interface was utilized to provide all necessary information through so called *topics* to the ADF. The provided inputs are mainly information about the ego-vehicle's state (e.g., position, orientation, velocity, acceleration, and yaw-rate) and information about the perceived ego-vehicle's environment in the form of object lists. Since in this case the focus is on the evaluation of the guidance of the vehicle, the perception-stack of the ADF is abstracted through an ideal sensor-model to reduce the complexity of the framework. Next to the input information given through the implemented interface, the ADF is in highly need of an HD map given in the Lanelet2 format. Utilizing the entity of this information, the ADF is capable of planning a driveable vehicle trajectory from which scalar control variables (longitudinal acceleration, steering angle) are derived. These variables are feed-back to *CarMaker* using a ROS-topic again and thus closing the feedback loop (cf. "Co-simulation framework" in Figure 2) of the simulation. As mentioned in the previous paragraph the so called "testmanager" (cf. Figure 2) is responsible for loading, execution, and evaluation of a defined set of scenarios. In the beginning the first scenario is loaded and the "testmanager" waits until the simulation is initialized successfully. Afterwards the defined target position is provided to the ADF. The evaluation of the ADF's behavior is started when the ego vehicle starts to move for the first time. Through the execution of the scenario, the evaluation is performed continuously and specific classification criteria (e.g., collision with other objects, leaving the drivable space, standing still for a longer time period) are supervised. If one of these criteria fails, the scenario is aborted and the test is evaluated is failed otherwise, the evaluation is continued until the desired target is reached. Besides the classification criteria, Key Performance Indicators (KPIs) are determined continuously. These KPIs are utilized to evaluate the resulting driving behavior performed by the ADF. The relevant KPIs for this survey are treated in the following paragraph. If a scenario is terminated, all results are stored, and the simulation is terminated. The process is repeated until all predefined scenarios have been ran.

---

[5] https://github.com/ika-rwth-aachen/RoadGeneration

[6] https://commonroad.in.tum.de/

**Evaluation of Simulations**

As mentioned in the previous paragraph, KPIs are used to evaluation the resulting behavior of the ADF. For this survey, a subset of all available KPIs is treated, rating the longitudinal acceleration and jerk, as well as the occurring lateral acceleration and jerk which correlate to the sense of comfort of the passengers. Moreover, the proportion of time the vehicle is stationary with respect to the total scenario time is evaluated. Lastly, the final distance to the desired target position is evaluated too. It applies to all KPIs, that a resulting value equal or greater than one can be assessed as good behavior. This is done by normalizing each KPI with a reference value that comes from literature or experiments. E.g., a lateral acceleration of $3\frac{m}{s^2}$ is used as an acceptable value according to ISO11270[7], so when the actual measured value is below that, the KPI will be greater than one. For KPIs like the longitudinal deceleration a velocity depended characteristic line and not a constant value is used to rate the performance. The values for longitudinal KPIs are taken from ISO15622[8] which describes tolerable parameter limits for ACC systems.

Table 2: KPIs used for the evaluation of a simulation run

| KPI Name | Evaluation of | Reference Value |
| --- | --- | --- |
| KPI_lon_acc | Longitudinal acceleration | variable with velocity |
| KPI_lon_decel | Longitudinal deceleration | variable with velocity |
| KPI_lon_neg_jerk | Negative longitudinal jerk | variable with velocity |
| KPI_lon_pos_jerk | Positive longitudinal jerk | variable with velocity |
| KPI_lat_acc | Lateral acceleration | $3\frac{m}{s^2}$ |
| KPI_lat_jerk | Lateral jerk | $5\frac{m}{s^3}$ |
| KPI_total_standstill_time | Vehicle standstill time | - |
| KPI_distance_target | Distance to the target location | $0,5\ m$ |

Based on these KPIs, experiments have been conducted to show that the proposed toolchain works meaningful. In the following sections, examples for variated parameters (cf. Table 1) of the two road networks shown in Figure 3 are presented and discussed.

*Example: Simple Road*

As shown in Table 1, six different logical experiments have been performed for the single road with a curve. Each variant has been sampled with ten concrete instances which results in 60 test runs in the described simulation setup for the first scenario.

The results of all KPIs for the varied radius of the curve are visualized in Figure 4 a) as a radar chart.

---

[7] https://www.iso.org/standard/50347.html (for LKA systems)

[8] https://www.iso.org/standard/71515.html (for ACC systems)

Simulative Performance Analysis of an AD Function with Road Network Variation

The curves consist of the mean of the minimal value for all 10 test runs for each set and correspond to the sets introduced in Table 1. It can be seen that most KPIs do not change significantly. However, the lateral acceleration behaves the best when the curve is the widest. Similar observations can be seen at the lateral jerk. When looking at the KPI for the longitudinal deceleration it is noticeable that the wider the curve, the better this KPI becomes. These results imply that the prototypical ADF behaves better (according to the KPIs) when curves are not too narrow.

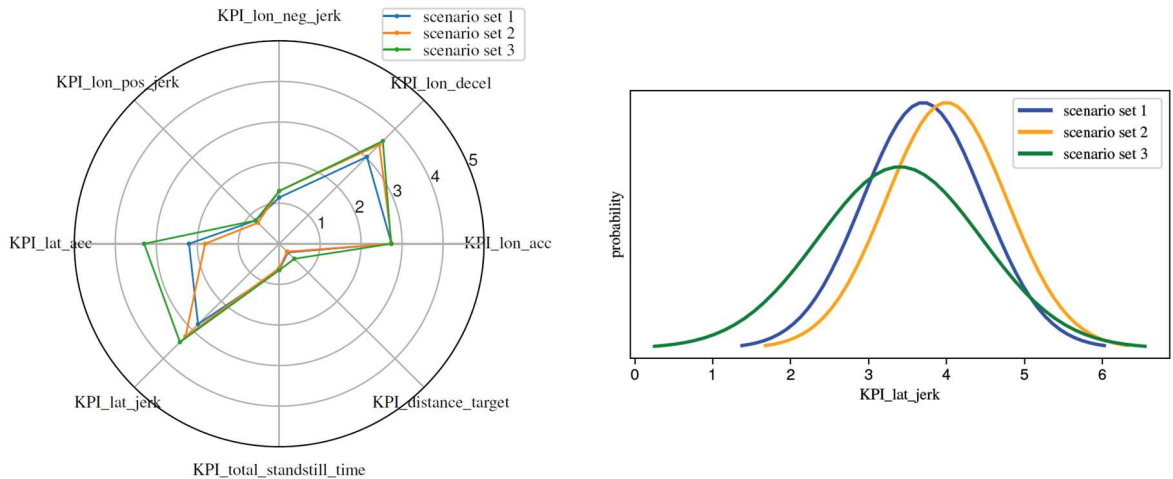

a) Radar chart of KPIs for different curve radii     b) Lateral jerk KPI for different lane widths

Figure 4: Radar charts of the experiments rolled out with the two parameter variations on a simple road.

The second experiment regards the lane width along the whole road of the driving lane. The mean value has been selected from 3.0m up to 3.75m as described in Table 1. Figure 4 b) shows the normal distribution of the lateral jerk KPI for each set. This representation is chosen since in a radar chart, similar to the curve radius variation, no differences are noticeable. It is noticeable that for the two smaller lane widths the KPI has better results than for the widest lane width. This may be explained by the fact that in a wider road the ADF has to adjust its lateral position more often to remain in the lane center since the room for deviations is more tolerable.

*Example: T-Junction*

In this section, the experiments using the T-junction are discussed. Again, Table 1 has been used to sample in total 60 concrete scenarios: for each of the two parameters three sets with 10 instances, respectively. Figure 5 shows radar charts of both parameters and the curves again correspond to the mean of the minimal occurring KPI values during the simulations.

When looking at the intersection angle (cf. Figure 5 a), the lateral jerk and lateral acceleration KPIs become better with increasing intersection angle, which indicates that the lateral performance of the vehicle is improving. The reason for this phenomenon might be that the larger the intersection angle of roads, the smoother the connecting road within the junction, so there is no violent lateral jerk when the ADF is driving on the connecting road. Another observation is that the longitudinal deceleration is better rated with decreasing intersection angles. This seems counterintuitive but might indicate that narrow curves can be better planned and result in smoother decelerations. The distance to target KPI has outliers which can be explained by the route planning algorithm that sometimes stop too early or overshoots the target destination.

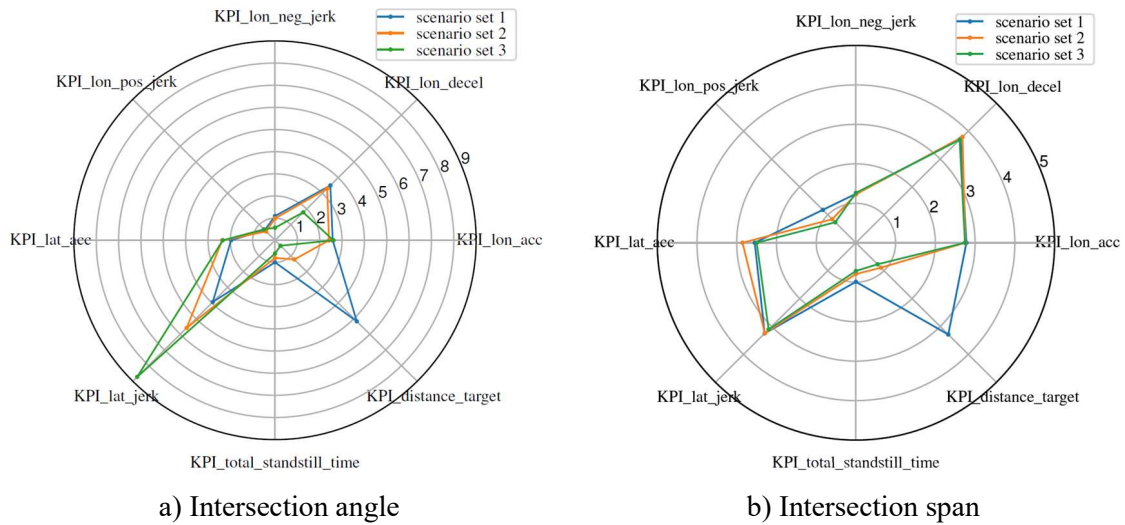

a) Intersection angle    b) Intersection span

Figure 5: Radar charts of the experiments rolled out with the two parameter variations on the T-junction.

The results of second experiment on the T-junction are shown in Figure 5 b) in the same manner as for the intersection angle. The results for all intersection sizes seem to be similar. A greater intersection span means that the turning lanes on the intersection are getting wider which leads to the thought that an ADF can handle wider turning lanes better than narrow curves. However, the utilized ADF seems to perform well on different intersection sizes. Again, the distance to target KPI has insufficient values smaller than one which could result from the same reason explained above.

**Conclusion & Outlook**

In conclusion, this paper presents an approach to systematically test a prototypical ROS based ADF with variations of road networks within the commercial simulation tool IPG *CarMaker*. The outcome is rated with a set of KPIs that can give an objective measure on how well the scenario has been executed by the ADF. The methodology has been applied to two rather straightforward road networks to demonstrate the proof of concept. The results show that in general the tool chain works and that the variated parameters partly have an influence on the performance of the ADF. The intersection angle for example influences the lateral KPIs. However, the size of the intersection seems to have no significant influence, but this must be validated more extensively. In general, the evaluation points out tendencies only and further parameters and parameter ranges should be investigated. Consequently, the approach will be further developed in the future.

The goal is to end up with a framework that is completely based on open-source tools and standard formats as interfaces. Thus, the next step is to formulate the scenarios in a standardized format as it is already done for the simulation maps and then be able to execute them in other simulation tools without the necessity of adapting them. At the current stage this has not been practicable since the described "testmanager" is based on IPG's in-house formats. Subsequently, the usage of the OpenSCENARIO standard for the whole scenario specification will be realized first. One tool that is becoming more and more popular is the open-source simulation software Carla and we will implement our test framework in there as a further step.

On the other hand, supplementary to the technical development, the complexity of the used road networks will be increased to further investigate the outlined hypothesis that road network elements can challenge an ADF. For now, rather straightforward simulation maps are used to show that the concept

Simulative Performance Analysis of an AD Function with Road Network Variation

works in principle. Consequently, the number of variated parameters per simulation will be extended and more complex junction designs will be generated. Further, road networks with more than one junction and connection road can be designed and scenarios where lane changes are required to reach the target destination are going to be created to test more complex maneuvers of the ADF.

**Acknowledgement**

This research is funded by the VVM project research initiative, promoted by the German Federal Ministry for Economic Affairs and Climate Action (BMWK).